\documentclass[aps, prb, a4paper, reprint, showkeys, nobibnotes, superscriptaddress]{revtex4-1}

\usepackage{graphicx}
\usepackage{bm}
\usepackage{amsmath}
\usepackage[amssymb]{SIunits}
\usepackage{nicefrac}
\usepackage{textcomp}
\usepackage{epstopdf}
\usepackage{hyperref}

\hypersetup{
    colorlinks=true,
    linkcolor=blue,
    citecolor=blue,
    filecolor=black,
    urlcolor=black,
}

\newcommand\mhyphen{{\operatorname{-}}}

\begin{document}

\title{Effects of Neutron Irradiation on Pinning Force Scaling in State-of-the-Art \\ {N}b$_3${S}n Wires}

\author{T.\ Baumgartner}
\email{tbaumgartner@ati.ac.at}
\author{M.\ Eisterer}
\author{H.\ W.\ Weber}
\affiliation{Atominstitut, Vienna University of Technology, Stadionallee 2, 1020 Vienna, Austria}

\author{R.\ Fl{\"u}kiger}
\author{C.\ Scheuerlein}
\author{L.\ Bottura}
\affiliation{CERN, 1211 Geneva 23, Switzerland}

\date{October 31, 2013}

\begin{abstract}

We present an extensive irradiation study involving five state-of-the-art Nb$_3$Sn wires which were subjected to sequential neutron irradiation up to a fast neutron fluence of $1.6 \cdot 10^{22}$\,m$^{-2}$ ($E > 0.1$\,MeV). The volume pinning force of short wire samples was assessed in the temperature range from 4.2 to 15\,K in applied fields of up to 7\,T by means of SQUID magnetometry in the unirradiated state and after each irradiation step. Pinning force scaling computations revealed that the exponents in the pinning force function differ significantly from those expected for pure grain boundary pinning, and that fast neutron irradiation causes a substantial change in the functional dependence of the volume pinning force. A model is presented, which describes the pinning force function of irradiated wires using a two-component ansatz involving a point-pinning contribution stemming from radiation induced pinning centers. The dependence of this point-pinning contribution on fast neutron fluence appears to be a universal function for all examined wire types.

\end{abstract}


\maketitle

\section{Introduction \label{sec:Intro}}

The fabrication of multifilamentary Nb$_3$Sn wires capable of carrying high critical current densities has advanced considerably in recent years. Various production techniques such as the Bronze Route, the Powder-In-Tube (PIT) process, and the Restacked Rod Process (RRP) are available, and superconductor geometries ranging from fine filaments with diameters of a few micrometers to large tubular sub-elements with outer diameters of roughly 100\,{\micro}m can be realized. It was discovered a long time ago that the properties of Nb$_3$Sn can be improved by adjusting the heat treatments which affect the grain morphology, as well as by introducing additives (usually Ta, Ti, or both) which increase the upper critical field. \cite{Luhman:Metallurgy} Today, due to long-standing efforts to improve this technologically highly relevant material, Nb$_3$Sn wires can be optimized using sophisticated production techniques and comprehensive knowledge of the underlying physics. \cite{Fluekiger:microstructure}

The effects of fast neutron irradiation (particle energy sufficient to displace atoms) on Nb$_3$Sn wires were studied extensively in the 1970s and 80s. In general an increase of the critical current density $J_{\text{c}}$ was observed at low fluences, followed by a decrease at high fluences. The resulting peak in the dependence of $J_{\text{c}}$ on fluence was usually found in the fast neutron fluence range of $2\,{\mhyphen}\,7 \cdot 10^{21}$\,m$^{-2}$. \cite{Snead:irrad_Nb3Sn, Hahn:simulation, Weiss:Nb3Sn_irrad, Hahn:fusion_neutron_irradiation} This behavior was often attributed to the increase of the upper critical field $B_{\text{c2}}$ due to the introduction of disorder, and the degradation of the critical temperature $T_{\text{c}}$ which dominates at high fluences. However, in some publications changes in the functional dependence of the volume pinning force were reported, which indicate the introduction of new pinning centers. \cite{Guinan:low_temp_irradiation, Okada:irrad} The results of many neutron irradiation studies as well as the concept of damage energy scaling, which facilitates the inference of effects caused by irradiation with a certain spectrum from data obtained using another spectrum, are discussed in a review article by Weber. \cite{Weber:radiation_effects}

Since state-of-the-art Nb$_3$Sn wires are significantly different from those available at the time of the cited irradiation studies, the investigation of their response to radiation damage is technologically relevant to projects such as particle accelerators and nuclear fusion devices. In particular, this knowledge is required for a life-time estimation of next generation quadrupole magnets which are envisaged for an upgrade of the Large Hadron Collider (LHC), where they will be exposed to a complex and intense radiation field near the interaction points. \cite{Fluekiger:RRP_PIT_irrad} Radiation induced changes in flux pinning and in the intrinsic properties of five candidate wires for this project were assessed by means of sequential irradiation of short wire samples in the TRIGA \mbox{Mark-II} reactor at the Atominstitut. It was found that the functional dependence of the volume pinning force of these wires is strongly affected by fast neutron irradiation, whereas the upper critical field increases only slightly. In the present work these changes in the flux pinning behavior are analyzed using the Unified Scaling Law. \cite{Ekin:USL} A model is presented, which describes the volume pinning force of irradiated wires based on their initial pinning properties and on a second mechanism stemming from radiation induced defects.

\section{Experimental \label{sec:Exp}}

\subsection{Samples \label{sec:Exp_Samples}}

Five types of state-of-the-art multifilamentary Nb$_3$Sn wires were examined in this work. Two of them are RRP wires containing Ta and Ti additions, respectively, two are PIT wires with Ta additions, and one is a binary (without any additives) prototype internal tin (IT) strand. All five wire types contain tubular sub-elements embedded in a copper matrix. \autoref{tab:sample_specs} lists the production process, the additive element, the number of sub-elements $N$, the inner and outer sub-element radii $\rho_{\text{i}}$ and $\rho_{\text{o}}$, and the critical temperature of each wire. The sub-element radii were obtained from scanning electron microscopy images using a procedure which counts pixels belonging to the \mbox{A-15} region of the sub-elements, thus yielding effective radii even for non-circular sub-element cross sections (with the exception of the two PIT wires, the cross sections are hexagonal in shape).

\begin{table}
	\caption{Characteristics of the five examined wire types. \label{tab:sample_specs}}
	\begin{ruledtabular}
		\begin{tabular}{ccccccc}
			\textbf{Name} & \textbf{Process} & \textbf{Add.} & $\bm{N}$ & $\bm{\rho_{\text{\textbf{i}}}}$ \text{({\micro}m)} & $\bm{\rho_{\text{\textbf{o}}}}$ \text{({\micro}m)} & $\bm{T_{\textbf{c}}}$ (K)\\
			\hline \\[-1.5ex]
			RRP-54  & RRP & Ta & 54  & 21.1 & 37.2 & 17.85\\
			PIT-192 & PIT & Ta & 192 & 12.4 & 22.5 & 17.64\\
			RRP-108 & RRP & Ti & 108 & 15.4 & 26.7 & 17.43\\
			IT-246 & IT & --- & 246 & 17.0 & 27.0 & 17.17\\
			PIT-114 & PIT & Ta & 114 & 11.7 & 20.5 & 17.66\\
		\end{tabular}
	\end{ruledtabular}
\end{table}

Straight wire pieces, a few centimeters in length, were heat treated at CERN after sealing in quartz glass tubes and then shipped to the Atominstitut. Roughly 4\,mm long samples for magnetization measurements were cut from the wires with a low speed diamond saw. This method is preferable to using a wire cutter, which might introduce mechanical damage in the sub-elements near the cut faces. Somewhat longer samples (about 2\,cm) for resistivity measurements were prepared with a wire cutter, since this measurement method is not sensitive to sub-element damage near the sample ends.

\subsection{Neutron irradiation \label{sec:Exp_Irrad}}

The sequential irradiation of 4\,mm long wire samples with fast neutrons was carried out in the central irradiation facility of the TRIGA Mark-II reactor at the Atominstitut. The fast neutron flux density $\varPhi_{\text{f}}$ at the sample position was approximately $4.5 \cdot 10^{16}$\,m$^{-2}$s$^{-1}$ ($E > 0.1$\,MeV) at the time of this irradiation program. The samples were sealed in quartz glass tubes to prevent direct contact with the reactor water, and irradiated at ambient reactor temperature (approx.\ 70\,{\textcelsius}). Once a proper decay time had passed in order to reduce exposure to radioactivity, the measurements described in \autoref{sec:Exp_SQUID} were carried out. Repeating this procedure several times with the same samples allowed a detailed assessment of changes in $J_{\text{c}}$ and $T_{\text{c}}$.

A small sample (roughly 5\,mg mass) of a high purity Ni foil was included in each irradiation step for fluence monitoring. The $^{58}\text{Ni}(\text{n},\text{p})^{58}\text{Co}$ reaction, whose threshold energy is roughly $1$\,MeV, was used to calculate the fast neutron fluence based on the activity of the radioactive isotope $^{58}\text{Co}$ which accumulates inside the monitor sample during irradiation. The maximum cumulative (including all irradiation steps) fast neutron fluence was $1.6 \cdot 10^{22}$\,m$^{-2}$. Since some wire types were included in the irradiation program later than others, not all of them reached the same cumulative fluence.

The presence of Ta in the wire types RRP-54, PIT-192, and PIT-114 leads to the formation of the radioactive isotope $^{182}$Ta, which has a half life of 115 days. Given the neutron spectrum of the reactor at the Atominstitut and the Ta content of these wires, a typical irradiation step of $\varPhi_{\text{f}} t = 2 \cdot 10^{21}$\,m$^{-2}$ leads to a $^{182}$Ta activity per wire length of roughly 100\,MBq$/$m shortly after irradiation. Therefore, magnetization measurements on short wire samples are preferable to transport current measurements in terms of radiation protection, since the latter require much longer pieces of wire.

\subsection{SQUID magnetometry \label{sec:Exp_SQUID}}

A Quantum Design MPMS~XL SQUID magnetometer was used to measure the magnetic moment of the 4\,mm long wire samples at different temperatures ranging from 4.2 to 15\,K. The samples were aligned with their axis perpendicular to the applied magnetic field. Magnetization loops were recorded with a step size of 0.2\,T up to an applied field of 7\,T, which is the maximum field the superconducting magnet inside the MPMS~XL can generate. The irreversible magnetic moment $m_{\text{irr}}$ as a function of applied field $B_{\text{a}}$ was obtained from the magnetization loops using
\begin{equation}
	m_{\text{irr}}(B_{\text{a}}) = \frac{1}{2} \, \Big( m_{\text{dec}}(B_{\text{a}}) - m_{\text{inc}}(B_{\text{a}}) \Big) \; \text{,}
	\label{eq:m_irr_loops}
\end{equation}
where $m_{\text{inc}}$ and $m_{\text{dec}}$ denote the magnetic moment measured in increasing and in decreasing field, respectively. The critical current density $J_{\text{c}}$ as a function of magnetic field was then evaluated at each temperature as described in \autoref{sec:Results_Fp}. This method allows the evaluation of $J_{\text{c}}$ in field and temperature ranges not easily accessible in transport current measurements due to the very high critical currents.

The same system was used to assess the critical temperature of the samples by means of AC susceptibility measurements, using a frequency of 30\,Hz and an amplitude of 30\,{\micro}T. The temperature at which the mid-point value of the superconducting to normal transition in the susceptibility curve occurs was taken as $T_{\text{c}}$. In \autoref{tab:sample_specs} the thus obtained critical temperature values before irradiation are listed for each examined wire type.

\subsection{Resistivity measurements \label{sec:Exp_Res}}

A cryostat equipped with a 15\,T superconducting magnet was used to perform resistivity measurements in order to obtain the upper critical field of the wires at various temperatures. An approximately 2\,cm long unirradiated sample of each wire type was mounted perpendicular to the applied field inside the VTI (variable temperature insert) of the cryostat. A current of 100\,mA was applied, and the voltage drop across the sample was measured using four-terminal sensing. By measuring the sample voltage at constant applied field $B_{\text{a}}$ while slowly changing the temperature inside the VTI, $T_{\text{c}}(B_{\text{a}})$ was determined from the change in the voltage associated with the transition between superconducting and normal state. This procedure was repeated at different applied fields, thus allowing the assessment of the temperature dependence of the upper critical field at $T \gtrsim T_{\text{c}} / 2$.

Resistivity measurements were also carried out with three irradiated short samples of different wire types, which were normally used for SQUID magnetometry. Satisfactory results were obtained from these samples in spite of their small length (approx.\ 4\,mm) which made it difficult to attach four contacts.

\section{Results and discussion \label{sec:Results}}

\subsection{Volume pinning force from magnetometry data \label{sec:Results_Fp}}

A $J_{\text{c}}$ evaluation model based on the Bean critical state model and on some geometric assumptions was used to obtain critical current densities from the magnetometry data. This model is described in a previous publication and was shown to yield results which are in satisfactory agreement with transport current measurements. \cite{Baumgartner:Jc_mag} According to this model, $J_{\text{c}}$ is related to $m_{\text{irr}}$ by
\begin{equation}
	J_{\text{c}} = \frac{3 m_{\text{irr}}}{4 N L \, ({\rho_{\text{o}}}^3 - {\rho_{\text{i}}}^3)} \; \text{,}
	\label{eq:Jc_eval}
\end{equation}
where $N$ is the number of sub-elements, $L$ is the sample length, and the inner and outer sub-element radii are denoted by $\rho_{\text{i}}$ and $\rho_{\text{o}}$, respectively. Note that the procedure for obtaining $m_{\text{irr}}$ described in \autoref{sec:Exp_SQUID} entails a first-order self-field correction, since the direction of the irreversible currents and thus the sign of the self-field changes when the applied field changes from increasing to decreasing field. An additional self-field correction was found to be unnecessary, since the associated changes in the field dependence of the critical current density would be insignificant. \cite{Baumgartner:PhD} Therefore, with the exception of very small applied fields, the equality $J_{\text{c}}(B) = J_{\text{c}}(B_{\text{a}})$ between the values of the critical current density at a certain field $B$ inside the superconducting sub-elements and at a certain applied field $B_{\text{a}}$ can be assumed.

The evaluation of $J_{\text{c}}(B)$ at different temperatures based on magnetometry data was performed on several (between four and six) unirradiated samples of each wire type, and on all samples included in each irradiation step (usually only one of each type at any particular fluence value). Data points which were dubious due to incomplete field penetration or because of flux jumps were excluded. Average values were calculated from the results obtained from unirradiated samples in order to have a reliable reference for investigating changes introduced by fast neutron irradiation. Barring data points measured at field values close to or above $B_{\text{c2}}$, the standard deviations of the $J_{\text{c}}(B)$ values were less than 5\% in all wire types.

The volume pinning force $F_{\text{p}} = \vert \vec{B} \times \vec{J_{\text{c}}} \vert$ can easily be obtained from the critical current density evaluated from  magnetometry data, since the applied field and the irreversible currents are in good approximation perpendicular to each other, as it is also assumed by the $J_{\text{c}}$ evaluation model. The cross product thus simplifies to $F_{\text{p}} = J_{\text{c}} \, B$. Note that the functional dependence of the volume pinning force is not sensitive to potential corrections to the purely geometric proportionality factor between $J_{\text{c}}$ and $m_{\text{irr}}$ in \autoref{eq:Jc_eval}.

\subsection{Pinning force scaling of unirradiated samples \label{sec:Results_Scaling_Unirr}}

Nb$_3$Sn is among the superconductors for which a scaling behavior of the volume pinning force was observed, i.e.\ coalescence into a single curve of normalized pinning force data recorded at different temperatures and plotted as a function of reduced field. The unified scaling law (USL) is a powerful tool for the analysis of pinning force scaling, and is described in detail in a review article by Ekin. \cite{Ekin:USL} In its separable form
\begin{equation}
	F_{\text{p}} = C \, g(\varepsilon) \, h(t) \, f(b)
	\label{eq:USL_separable}
\end{equation}
it expresses the volume pinning force $F_{\text{p}}$ as a product of a constant $C$ and three mutually independent functions $g(\varepsilon)$, $h(t)$, and $f(b)$, which describe the dependence on strain, temperature, and magnetic field, respectively. The field dependence $f(b)$, which is often referred to as the pinning force function, is cast in the form
\begin{equation}
	f(b) = b^p \, (1 - b)^q
	\label{eq:f}
\end{equation}
within the framework of the USL. In \autoref{eq:f} $p$ and $q$ are the so-called low- and high-field exponents, and $b$ is the reduced magnetic field, which is defined as
\begin{equation}
	b = \frac{B}{B_{\text{c2}}^*(t, \varepsilon)} \; \text{.}
	\label{eq:b_USL}
\end{equation}
$B_{\text{c2}}^*(t, \varepsilon)$ is a temperature and strain dependent scaling field, which is in general close to but not necessarily identical to the upper critical field $B_{\text{c2}}$. Since the former is an effective depinning field which is commonly used as a fit parameter, it is determined by the functional dependence of the volume pinning force, whereas the latter marks the thermodynamic transition between superconducting and normal phase.

The function $f(b)$ was computed for each wire type in its unirradiated state using the average $J_{\text{c}}(B,T)$ values obtained from SQUID magnetometry. An algorithm specifically designed for this purpose normalizes the $F_{\text{p}}(B)$ data, and finds the optimum values of $p$ and $q$ by minimizing the global error including data obtained at different temperatures, using $B_{\text{c2}}^*(T)$ as a fit parameter.

\begin{figure}
	\includegraphics{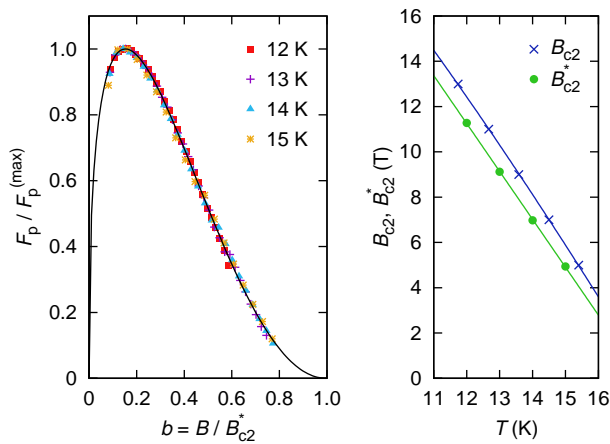}%
	\caption{Normalized pinning force data obtained from magnetometry, pinning force function $f(b)$, and comparison of the temperature dependences of the upper critical field and the scaling field of the RRP-54 wire. \label{fig:scal_RRP_unirr}}
\end{figure}

\begin{figure}
	\includegraphics{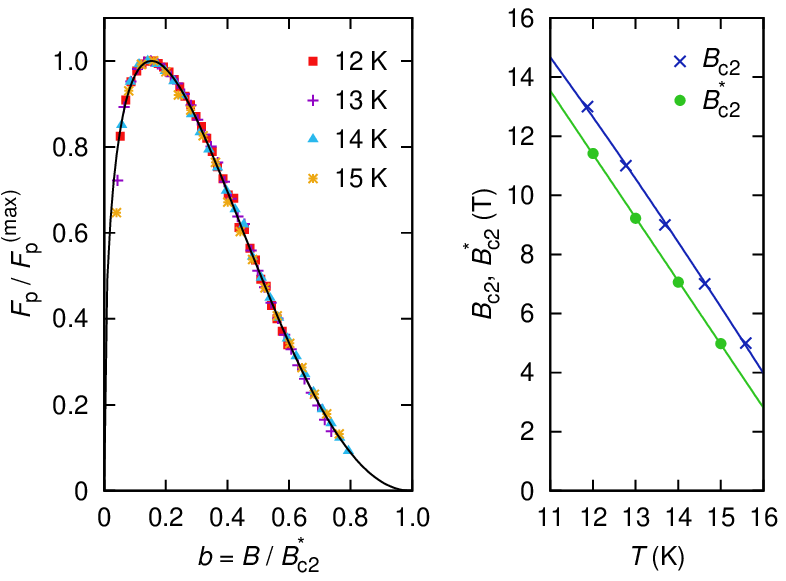}%
	\caption{Magnetometry data, $f(b)$, and comparison of $B_{\text{c2}}(T)$ and $B_{\text{c2}}^*(T)$ of the PIT-192 wire. \label{fig:scal_PIT_unirr}}
\end{figure}

\begin{figure}
	\includegraphics{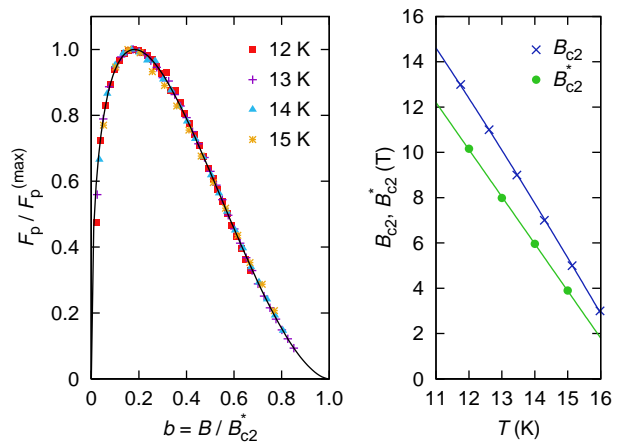}%
	\caption{Magnetometry data, $f(b)$, and comparison of $B_{\text{c2}}(T)$ and $B_{\text{c2}}^*(T)$ of the RRP-108 wire. \label{fig:scal_TiR_unirr}}
\end{figure}

\begin{figure}
	\includegraphics{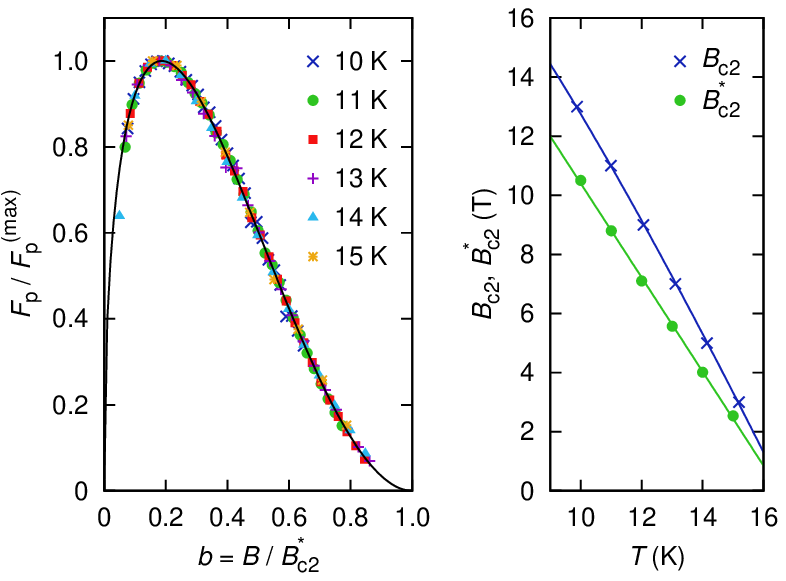}%
	\caption{Magnetometry data, $f(b)$, and comparison of $B_{\text{c2}}(T)$ and $B_{\text{c2}}^*(T)$ of the IT-246 wire. \label{fig:scal_BIN_unirr}}
\end{figure}

\begin{figure}
	\includegraphics{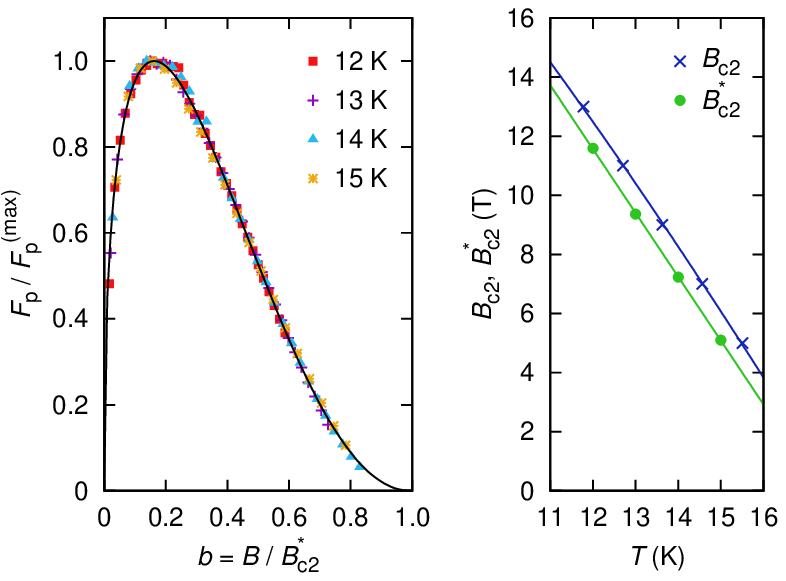}%
	\caption{Magnetometry data, $f(b)$, and comparison of $B_{\text{c2}}(T)$ and $B_{\text{c2}}^*(T)$ of the PIT-114 wire. \label{fig:scal_P07_unirr}}
\end{figure}

The quality of the thus computed pinning force functions suffers if the magnetometry data do not cover a sufficiently wide range of reduced field values. Therefore, only data obtained at temperatures where 7\,T (maximum field of the magnetometer) exceeds $B_{\text{c2}} / 2$ were included. The resulting $f(b)$ curves (normalized to a peak value of 1) as well as a comparison of the temperature dependences of $B_{\text{c2}}^*$ and $B_{\text{c2}}$ are shown in \autoref{fig:scal_RRP_unirr} for the RRP-54 wire, in \autoref{fig:scal_PIT_unirr} for the PIT-192 wire, in \autoref{fig:scal_TiR_unirr} for the RRP-108 wire, in \autoref{fig:scal_BIN_unirr} for the IT-246 wire, and in \autoref{fig:scal_P07_unirr} for the PIT-114 wire.

\autoref{tab:scal_para_unirr} lists the values of $p$ and $q$ as well as the value $b_{\text{max}} = p / (p + q)$ of the reduced field at which the function $f(b)$ peaks, and the temperature ranges the scaling computations are based on. In the case of pure grain boundary pinning, which is generally assumed to be the primary pinning mechanism in Nb$_3$Sn, the scaling exponents are expected to be $p = \nicefrac{1}{2}$ and $q = 2$, corresponding to $b_{\text{max}} = 0.2$. \cite{Dew-Hughes:flux_pinning} It is noteworthy that the exponents obtained from the binary wire IT-246 are closest to the expected values, whereas the ternary wires exhibit significant deviations from these values.

\begin{table}
	\caption{Scaling exponents of the unirradiated wire samples, resulting peak positions, and temperature ranges used for the scaling computations ($7\,\text{T} > B_{\text{c2}} / 2$). \label{tab:scal_para_unirr}}
	\begin{ruledtabular}
		\begin{tabular}{ccccc}
			\textbf{Wire type} & $\bm{p}$ & $\bm{q}$ & $\bm{b_{\textbf{max}}}$ & $\bm{T}$ \textbf{range} (K) \\
			\hline \\[-1.5ex]
			RRP-54  & 0.379 & 2.102 & 0.153 & 12 -- 15 \\
			PIT-192 & 0.383 & 2.117 & 0.153 & 12 -- 15 \\
			RRP-108 & 0.384 & 1.737 & 0.181 & 12 -- 15 \\
			IT-246  & 0.447 & 1.942 & 0.187 & 10 -- 15 \\
			PIT-114 & 0.417 & 2.149 & 0.163 & 12 -- 15 \\
		\end{tabular}
	\end{ruledtabular}
\end{table}

Fl{\"u}kiger et al.\ attributed the deviations of the scaling exponents from $p = \nicefrac{1}{2}$ and $q = 2$ in Bronze Route wires to the fact that the Nb$_3$Sn in these wires is comprised of two grain types which differ significantly in geometry and Sn content (equiaxed and columnar grains). \cite{Fluekiger:optimization} It stands to reason that the $p$ and $q$ values in \autoref{tab:scal_para_unirr} can also be explained based on the grain morphology of the examined wires. The two PIT wires possess a dual grain morphology, which is reflected by a double-transition when $T_{\text{c}}$ is measured using the AC susceptibility method, whereas the responsible morphological features may be more subtle in the other wire types. Element concentration profiles obtained from scanning transmission electron microscopy suggest that grains in ternary Nb$_3$Sn wires form thicker boundaries, which may explain why the wires containing additives deviate more from the theoretical scaling exponents than the binary wire. \cite{Rodrigues:grain_boundaries}

\subsection{Effects of irradiation on pinning force scaling \label{sec:Results_Scaling_Irrad}}

When the pinning force data obtained from magnetization measurements on irradiated samples were analyzed using the above-mentioned algorithm, significant changes in the scaling exponents $p$ and $q$ relative to the unirradiated state were found, whereas the scaling field $B_{\text{c2}}^*$ was barely affected. A closer examination of the data revealed that the changes in $p$ and $q$ were the result of a shift of the peak in the pinning force function towards higher values of $b$. Such a shift was previously reported by Guinan et al.\ as well as by Okada et al.\ in Nb$_3$Sn wires irradiated with 14.8\,MeV neutrons and fast reactor neutrons, respectively. \cite{Guinan:low_temp_irradiation, Okada:irrad} Similar changes in the functional dependence of the volume pinning force were found by Seibt in V$_3$Ga irradiated with 50\,MeV deuterons. \cite{Seibt:deuteron_irrad}

Since this change in the shape of $f(b)$ indicates the presence of radiation induced pinning centers, an algorithm was devised which assumes that the pinning force function can be modeled as the sum of two contributions with the relative weights $\alpha$ and $\beta$, where $\alpha + \beta = 1$:
\begin{equation}
	f(b) = \alpha \, b^{p_1} \, (1 - b)^{q_1} + \beta \, b^{p_2} \, (1 - b)^{q_2}
	\label{eq:f_2M}
\end{equation}
The exponents $p_1$ and $q_1$ were fixed to the values obtained from measurements on unirradiated samples (cf.\ \autoref{tab:scal_para_unirr}), and several combinations of $p_2$ and $q_2$, which were calculated by Dew-Hughes for different pinning mechanisms, were tested. \cite{Dew-Hughes:flux_pinning} The exponents $p_2 = 1$ and $q_2 = 2$, which correspond to pinning by core interaction between the flux lines and normal conducting point-like (small compared to the flux line spacing) structures, exhibited significantly better agreement with the data than any other set of values. The scaling field $B_{\text{c2}}^*(T)$ was used as a fit parameter, and the relative strength $\beta$ of the radiation induced pinning was found iteratively by computing the global deviation within the entire temperature range of interest (the same as listed in \autoref{tab:scal_para_unirr}). The idea of using a two-component ansatz to describe the pinning force in irradiated superconductors was already published by K{\"u}pfer et al.\ as well as by Maier and Seibt, although not explicitly in the form used in \autoref{eq:f_2M}. \cite{Kuepfer:Jc_irrad, Maier:irradiated_Nb3Sn}

\begin{figure}
	\includegraphics{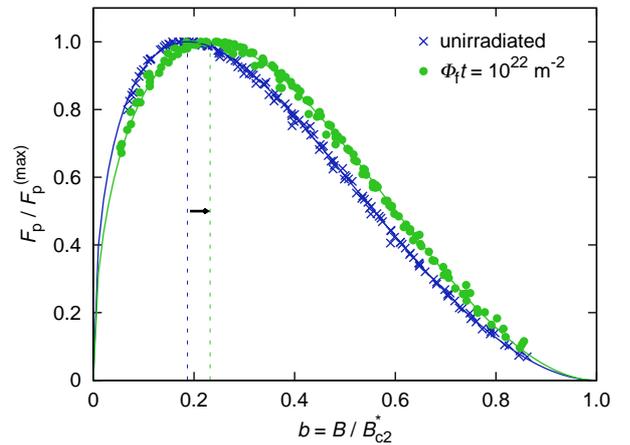}%
	\caption{Comparison of the pinning force scaling behavior (10\,K $\leq T \leq$ 15\,K) of the IT-246 wire in its unirradiated state and after irradiation to $\varPhi_{\text{f}} t = 10^{22}$\,m$^{-2}$. A shift of the peak position towards higher values of $b$ is clearly visible. \label{fig:peak_shift_BIN}}
\end{figure}

As an example, \autoref{fig:peak_shift_BIN} shows the thus calculated two-mechanism $f(b)$ function (including a normalizing factor) and the normalized pinning force data of the IT-246 wire after irradiation to a fast neutron fluence of $10^{22}$\,m$^{-2}$, as well as the pinning force function found in the unirradiated state for comparison. The peak position shifts to $b_{\text{max}} = 0.232$, which equates to a 24\% change relative to the unirradiated state, and the corresponding value of $\beta$ is 0.449. In the extreme case of $\beta = 1$ (original pinning mechanism insignificant compared to the point-pinning contribution) the position of the peak would be at $b_{\text{max}} = \nicefrac{1}{3}$.

\begin{figure}
	\includegraphics{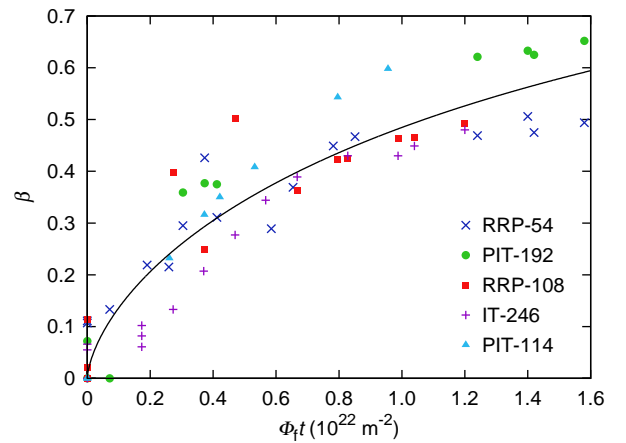}%
	\caption{Point-pinning contribution $\beta$ of the five examined wire types as a function of fast neutron fluence. \label{fig:beta}}
\end{figure}

Pinning force data obtained from all five wire types after each irradiation step were analyzed using this two-mechanism algorithm. The resulting dependence of the point-pinning contribution $\beta$ on fast neutron fluence is shown in \autoref{fig:beta}. Although there is considerable scatter in the plot, a clear trend is discernible, which appears to be universal for all examined wire types: At low fluences $\beta$ increases steeply, whereas it seems to exhibit a saturation behavior at high fluences. The function
\begin{equation}
	\beta(\varPhi_{\text{f}} \, t) = 1 - \text{e}^{- {\left( \varPhi_{\text{f}} \, t / \varPhi_{\text{f}} \, t^* \right)}^n}
	\label{eq:beta_fit}
\end{equation}
was found to be in general agreement with the observed behavior, and is shown as a solid line in \autoref{fig:beta}. The characteristic fast neutron fluence $\varPhi_{\text{f}} \, t^*$ and the exponent $n$ were fit to the data, yielding $\varPhi_{\text{f}} \, t^* = 1.87 \cdot 10^{22}$\,m$^{-2}$ and $n = 0.656$.

\begin{figure}
	\includegraphics{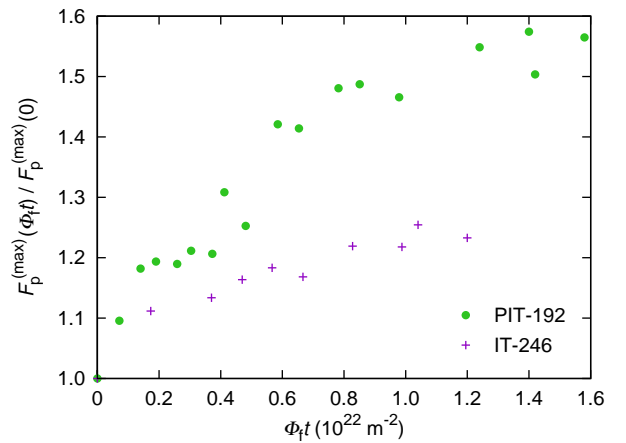}%
	\caption{Relative change of the maximum volume pinning force of the PIT-192 and the IT-246 wire types at $T = 4.2$\,K as a function of fast neutron fluence. \label{fig:Fp_max}}
\end{figure}

At $\beta \approx 0.5$, which occurs at a fast neutron fluence of roughly $10^{22}$\,m$^{-2}$, the contribution of the radiation induced pinning centers is approximately equal to the volume pinning force in the unirradiated state. This is consistent with calculations by F{\"a}hnle, who estimated the volume pinning force of  such pinning centers in Nb$_3$Sn at $\varPhi_{\text{f}} \, t = 10^{22}$\,m$^{-2}$, assuming that they are normal conducting regions with an average diameter of 10\,nm. The result was that the pinning force exerted by the radiation induced defects is comparable to typical values of $F_{\text{p}}$ in Nb$_3$Sn. \cite{Faehnle:influence_1} \autoref{fig:Fp_max} shows the relative change of the peak value of the volume pinning force of the PIT-192 and the IT-246 wire types as a function of fast neutron fluence. The plot contains SQUID magnetometry data recorded at $T = 4.2$\,K, since in the temperature range the scaling analysis is based on the absolute value of the maximum pinning force $F_{\text{p}}^{\text{(max)}}$ is influenced appreciably by the $T_{\text{c}}$ degradation. The latter amounts to approximately 3\% within the examined fluence range, as previously published. \cite{Fluekiger:RRP_PIT_irrad}

It is noteworthy that the volume pinning force in the PIT-192 wire increases by roughly 50\% at $\varPhi_{\text{f}} \, t = 10^{22}$\,m$^{-2}$ (the other alloyed wires exhibit a similar behavior), whereas the increase reaches only half of this value in the binary IT-246 wire, although the absolute values of $F_{\text{p}}^{\text{(max)}}$ in the unirradiated state are similar. \autoref{fig:beta} does not reflect this difference in the increase of the maximum volume pinning force at $T = 4.2$\,K due to deviations from the described scaling behavior at low temperatures. \cite{Baumgartner:PhD} The cause of the difference is currently unknown. Perhaps, for yet unknown reasons, the radiation induced defects have a higher pinning efficiency in ternary Nb$_3$Sn than in binary material. Such behavior might arise from differences in the migration of radiation induced defects during irradiation, which could be caused by the different grain boundary properties of binary and ternary Nb$_3$Sn mentioned in \autoref{sec:Results_Scaling_Unirr}. The radiation induced pinning centers responsible for the increase of the volume pinning force are obviously stable at room temperature, and earlier studies suggest that this stability is maintained up to much higher temperatures. \cite{Bett:irrad, Meier-Hirmer:V3Si_1}

\subsection{Effects of irradiation on the upper critical field \label{sec:Results_Bc2}}

\begin{figure}
	\includegraphics{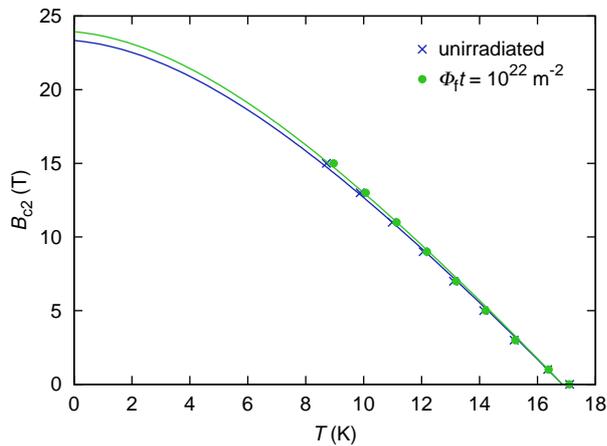}%
	\caption{Temperature dependence of the upper critical field of the IT-246 wire in its unirradiated state and after irradiation to $\varPhi_{\text{f}} t = 10^{22}$\,m$^{-2}$. The zero-temperature value exhibits only a slight increase of 2.6\%. \label{fig:Bc2_irrad_BIN}}
\end{figure}

As discussed in \autoref{sec:Exp_Res}, resistivity measurements were performed on unirradiated samples in order to assess the temperature dependence of the upper critical field of the wires. This procedure was also carried out with three irradiated samples, which facilitated a comparison of the changes in $B_{\text{c2}}$ with the insignificant increase in $B_{\text{c2}}^*$ found in the scaling computations. \autoref{fig:Bc2_irrad_BIN} shows a comparison of data obtained from an unirradiated IT-246 sample and from a sample of the same wire after irradiation to a fast neutron fluence of $10^{22}$\,m$^{-2}$. The extrapolations to low temperatures were computed using the function
\begin{equation}
	h_{\text{fit}}^*(t) = 1 - t - C_1 \, (1 - t)^2 - C_2 \, (1 - t)^4 \; \text{,}
	\label{eq:WHH_fit}
\end{equation}
with $C_1 = 0.153$ and $C_2 = 0.152$, which was found by the authors to be a good approximation to the dirty limit temperature dependence of the upper critical field calculated by Helfand and Werthamer. \cite{Baumgartner:PhD, WHH:Hc2_II} Using \autoref{eq:WHH_fit}, the upper critical field at arbitrary temperatures below $T_{\text{c}}$ is given by
\begin{equation}
	B_{\text{c2}}(T) = \frac{B_{\text{c2}}(0)}{0.693} \, h_{\text{fit}}^*(T / T_{\text{c}}) \; \text{.}
	\label{eq:Bc2_fit}
\end{equation}
The zero-temperature upper critical field $B_{\text{c2}}(0)$ and the critical temperature $T_{\text{c}}$ were used as fit parameters to obtain the curves shown in \autoref{fig:Bc2_irrad_BIN}. The fast neutron fluences the three irradiated samples were exposed to, their zero-temperature upper critical fields before and after irradiation, and the relative changes of the $B_{\text{c2}}(0)$ values are given in \autoref{tab:Bc2_change}. The three wire types exhibit only a minor radiation induced increase of the upper critical field, in accordance with the results of the scaling computations. This behavior is to be expected in wires whose upper critical field was optimized by the addition of Ta or Ti, since their initial normal state resistivity $\rho_{\text{n}}$ is high, and cannot be increased significantly by the introduction of additional disorder. \cite{Fluekiger:microstructure}

\begin{table}
	\caption{Comparison of the zero-temperature $B_{\text{c2}}$ values of three wire types in the unirradiated state and after irradiation to the specified fast neutron fluences. \label{tab:Bc2_change}}
	\begin{ruledtabular}
		\begin{tabular}{ccccc}
			\textbf{Wire type} & $\bm{\varPhi_{\textbf{f}} t}$ (m$^{-2}$) & $\bm{B^{\textbf{unirr}}_{\textbf{c2}}}$ (T) & $\bm{B^{\textbf{irrad}}_{\textbf{c2}}}$ (T) & \textbf{change} \\
			\hline \\[-1.5ex]
			RRP-108 & $1.0 \cdot 10^{22}$ & 29.72 & 30.25 & +1.8\% \\
			IT-246 & $1.0 \cdot 10^{22}$ & 23.34 & 23.93 & +2.6\% \\
			PIT-114 & $0.8 \cdot 10^{22}$ & 27.40 & 28.63 & +4.5\% \\
		\end{tabular}
	\end{ruledtabular}
\end{table}

In the case of the binary wire, however, the small $B_{\text{c2}}$ increase was somewhat surprising. For the lack of additives, its $\rho_{\text{n}}$ value in the unirradiated state should be lower than that of the ternary wires, leaving a larger relative increase of the upper critical field to be expected. An explanation for this discrepancy may lie in the fact that the Nb$_3$Sn in the IT-246 wire is fairly off-stoichiometric, as indicated by its relatively low $T_{\text{c}}$ of 17.17\,K (the lowest value among the five examined wire types), which leads to both a low $B_{\text{c2}}$ value and a high normal state resistivity. \cite{Fluekiger:microstructure}

\section{Conclusions \label{sec:Conclusions}}

A pinning force scaling analysis on SQUID magnetometry data obtained from five state-of-the-art Nb$_3$Sn wires revealed that their low- and high-field scaling exponents deviate significantly from the values expected for pure grain boundary pinning ($p = \nicefrac{1}{2}$, $q = 2$). The scaling exponents of the binary wire are closest to these values, whereas the four alloyed wires exhibit larger deviations. Differences in grain boundary thickness and composition may offer an explanation for this behavior.

Fast neutron irradiation up to a fluence of $1.6 \cdot 10^{22}$\,m$^{-2}$ was shown to have a pronounced effect on both the magnitude and the functional dependence of the volume pinning force of the examined wires. The pinning force function of irradiated wires was found to be in good agreement with a two-component model which adds the contribution of a radiation induced pinning mechanism to the original pinning force function. This second mechanism was identified as pinning by core interaction between flux lines and normal conducting point-like structures. It was shown by means of sequential irradiation that the dependence of the point-pinning contribution on fast neutron fluence appears to be a universal function for all examined wire types. Resistivity measurements demonstrated that the upper critical field increases only slightly after irradiation, which is in agreement with the results of the scaling analysis.

\bibliography{Bibliography}

\begin{thebibliography}{23}%
\makeatletter
\providecommand \@ifxundefined [1]{%
 \@ifx{#1\undefined}
}%
\providecommand \@ifnum [1]{%
 \ifnum #1\expandafter \@firstoftwo
 \else \expandafter \@secondoftwo
 \fi
}%
\providecommand \@ifx [1]{%
 \ifx #1\expandafter \@firstoftwo
 \else \expandafter \@secondoftwo
 \fi
}%
\providecommand \natexlab [1]{#1}%
\providecommand \enquote  [1]{``#1''}%
\providecommand \bibnamefont  [1]{#1}%
\providecommand \bibfnamefont [1]{#1}%
\providecommand \citenamefont [1]{#1}%
\providecommand \href@noop [0]{\@secondoftwo}%
\providecommand \href [0]{\begingroup \@sanitize@url \@href}%
\providecommand \@href[1]{\@@startlink{#1}\@@href}%
\providecommand \@@href[1]{\endgroup#1\@@endlink}%
\providecommand \@sanitize@url [0]{\catcode `\\12\catcode `\$12\catcode
  `\&12\catcode `\#12\catcode `\^12\catcode `\_12\catcode `\%12\relax}%
\providecommand \@@startlink[1]{}%
\providecommand \@@endlink[0]{}%
\providecommand \url  [0]{\begingroup\@sanitize@url \@url }%
\providecommand \@url [1]{\endgroup\@href {#1}{\urlprefix }}%
\providecommand \urlprefix  [0]{URL }%
\providecommand \Eprint [0]{\href }%
\providecommand \doibase [0]{http://dx.doi.org/}%
\providecommand \selectlanguage [0]{\@gobble}%
\providecommand \bibinfo  [0]{\@secondoftwo}%
\providecommand \bibfield  [0]{\@secondoftwo}%
\providecommand \translation [1]{[#1]}%
\providecommand \BibitemOpen [0]{}%
\providecommand \bibitemStop [0]{}%
\providecommand \bibitemNoStop [0]{.\EOS\space}%
\providecommand \EOS [0]{\spacefactor3000\relax}%
\providecommand \BibitemShut  [1]{\csname bibitem#1\endcsname}%
\let\auto@bib@innerbib\@empty
\bibitem [{\citenamefont {Luhman}\ and\ \citenamefont
  {Dew-Hughes}(1979)}]{Luhman:Metallurgy}%
  \BibitemOpen
  \bibinfo {editor} {\bibfnamefont {T.}~\bibnamefont {Luhman}}\ and\ \bibinfo
  {editor} {\bibfnamefont {D.}~\bibnamefont {Dew-Hughes}},\ eds.,\ \href@noop
  {} {\emph {\bibinfo {title} {Treatise on Materials Science and Technology,
  Volume 14: Metallurgy of Superconducting Materials}}}\ (\bibinfo  {publisher}
  {Academic Press, New York},\ \bibinfo {year} {1979})\BibitemShut {NoStop}%
\bibitem [{\citenamefont {Fl{\"u}kiger}\ \emph
  {et~al.}(2008{\natexlab{a}})\citenamefont {Fl{\"u}kiger}, \citenamefont
  {Uglietti}, \citenamefont {Senatore},\ and\ \citenamefont
  {Buta}}]{Fluekiger:microstructure}%
  \BibitemOpen
  \bibfield  {author} {\bibinfo {author} {\bibfnamefont {R.}~\bibnamefont
  {Fl{\"u}kiger}}, \bibinfo {author} {\bibfnamefont {D.}~\bibnamefont
  {Uglietti}}, \bibinfo {author} {\bibfnamefont {C.}~\bibnamefont {Senatore}},
  \ and\ \bibinfo {author} {\bibfnamefont {F.}~\bibnamefont {Buta}},\
  }\href@noop {} {\bibfield  {journal} {\bibinfo  {journal} {Cryogenics}\
  }\textbf {\bibinfo {volume} {48}},\ \bibinfo {pages} {293} (\bibinfo {year}
  {2008}{\natexlab{a}})}\BibitemShut {NoStop}%
\bibitem [{\citenamefont {Snead~{Jr.}}\ and\ \citenamefont
  {Parkin}(1976)}]{Snead:irrad_Nb3Sn}%
  \BibitemOpen
  \bibfield  {author} {\bibinfo {author} {\bibfnamefont {C.~L.}\ \bibnamefont
  {Snead~{Jr.}}}\ and\ \bibinfo {author} {\bibfnamefont {D.~M.}\ \bibnamefont
  {Parkin}},\ }\href@noop {} {\bibfield  {journal} {\bibinfo  {journal}
  {Nuclear Technology}\ }\textbf {\bibinfo {volume} {29}},\ \bibinfo {pages}
  {264} (\bibinfo {year} {1976})}\BibitemShut {NoStop}%
\bibitem [{\citenamefont {Hahn}\ \emph {et~al.}(1986)\citenamefont {Hahn},
  \citenamefont {Hoch}, \citenamefont {Weber}, \citenamefont {Birtcher},\ and\
  \citenamefont {Brown}}]{Hahn:simulation}%
  \BibitemOpen
  \bibfield  {author} {\bibinfo {author} {\bibfnamefont {P.~A.}\ \bibnamefont
  {Hahn}}, \bibinfo {author} {\bibfnamefont {H.}~\bibnamefont {Hoch}}, \bibinfo
  {author} {\bibfnamefont {H.~W.}\ \bibnamefont {Weber}}, \bibinfo {author}
  {\bibfnamefont {R.~C.}\ \bibnamefont {Birtcher}}, \ and\ \bibinfo {author}
  {\bibfnamefont {B.~S.}\ \bibnamefont {Brown}},\ }\href@noop {} {\bibfield
  {journal} {\bibinfo  {journal} {Journal of Nuclear Materials}\ }\textbf
  {\bibinfo {volume} {141--143}},\ \bibinfo {pages} {405} (\bibinfo {year}
  {1986})}\BibitemShut {NoStop}%
\bibitem [{\citenamefont {Weiss}\ \emph {et~al.}(1987)\citenamefont {Weiss},
  \citenamefont {Fl{\"u}kiger}, \citenamefont {Maurer}, \citenamefont {Hahn},\
  and\ \citenamefont {Guinan}}]{Weiss:Nb3Sn_irrad}%
  \BibitemOpen
  \bibfield  {author} {\bibinfo {author} {\bibfnamefont {F.}~\bibnamefont
  {Weiss}}, \bibinfo {author} {\bibfnamefont {R.}~\bibnamefont {Fl{\"u}kiger}},
  \bibinfo {author} {\bibfnamefont {W.}~\bibnamefont {Maurer}}, \bibinfo
  {author} {\bibfnamefont {P.~A.}\ \bibnamefont {Hahn}}, \ and\ \bibinfo
  {author} {\bibfnamefont {M.~W.}\ \bibnamefont {Guinan}},\ }\href@noop {}
  {\bibfield  {journal} {\bibinfo  {journal} {IEEE Transactions on Magnetics}\
  }\textbf {\bibinfo {volume} {23}},\ \bibinfo {pages} {976} (\bibinfo {year}
  {1987})}\BibitemShut {NoStop}%
\bibitem [{\citenamefont {Hahn}\ \emph {et~al.}(1991)\citenamefont {Hahn},
  \citenamefont {Guinan}, \citenamefont {Summers}, \citenamefont {Okada},\ and\
  \citenamefont {Smathers}}]{Hahn:fusion_neutron_irradiation}%
  \BibitemOpen
  \bibfield  {author} {\bibinfo {author} {\bibfnamefont {P.~A.}\ \bibnamefont
  {Hahn}}, \bibinfo {author} {\bibfnamefont {M.~W.}\ \bibnamefont {Guinan}},
  \bibinfo {author} {\bibfnamefont {L.~T.}\ \bibnamefont {Summers}}, \bibinfo
  {author} {\bibfnamefont {T.}~\bibnamefont {Okada}}, \ and\ \bibinfo {author}
  {\bibfnamefont {D.~B.}\ \bibnamefont {Smathers}},\ }\href@noop {} {\bibfield
  {journal} {\bibinfo  {journal} {Journal of Nuclear Materials}\ }\textbf
  {\bibinfo {volume} {179--181}},\ \bibinfo {pages} {1127} (\bibinfo {year}
  {1991})}\BibitemShut {NoStop}%
\bibitem [{\citenamefont {Guinan}\ \emph {et~al.}(1984)\citenamefont {Guinan},
  \citenamefont {{van Konynenburg}},\ and\ \citenamefont
  {Mitchell}}]{Guinan:low_temp_irradiation}%
  \BibitemOpen
  \bibfield  {author} {\bibinfo {author} {\bibfnamefont {M.~W.}\ \bibnamefont
  {Guinan}}, \bibinfo {author} {\bibfnamefont {R.~A.}\ \bibnamefont {{van
  Konynenburg}}}, \ and\ \bibinfo {author} {\bibfnamefont {J.~B.}\ \bibnamefont
  {Mitchell}},\ }\href@noop {} {\enquote {\bibinfo {title} {Effects of
  low-temperature fusion neutron irradiation on critical properties of a
  monofilament niobium-tin superconductor},}\ }\bibinfo {howpublished}
  {Informal Report, Lawrence Livermore National Laboratory} (\bibinfo {year}
  {1984})\BibitemShut {NoStop}%
\bibitem [{\citenamefont {Okada}\ \emph {et~al.}(1988)\citenamefont {Okada},
  \citenamefont {Fukumoto}, \citenamefont {Katagiri}, \citenamefont {Saito},
  \citenamefont {Kodaka},\ and\ \citenamefont {Yoshida}}]{Okada:irrad}%
  \BibitemOpen
  \bibfield  {author} {\bibinfo {author} {\bibfnamefont {T.}~\bibnamefont
  {Okada}}, \bibinfo {author} {\bibfnamefont {M.}~\bibnamefont {Fukumoto}},
  \bibinfo {author} {\bibfnamefont {K.}~\bibnamefont {Katagiri}}, \bibinfo
  {author} {\bibfnamefont {K.}~\bibnamefont {Saito}}, \bibinfo {author}
  {\bibfnamefont {H.}~\bibnamefont {Kodaka}}, \ and\ \bibinfo {author}
  {\bibfnamefont {H.}~\bibnamefont {Yoshida}},\ }\href@noop {} {\bibfield
  {journal} {\bibinfo  {journal} {Journal of Applied Physics}\ }\textbf
  {\bibinfo {volume} {63}},\ \bibinfo {pages} {4580} (\bibinfo {year}
  {1988})}\BibitemShut {NoStop}%
\bibitem [{\citenamefont {Weber}(2011)}]{Weber:radiation_effects}%
  \BibitemOpen
  \bibfield  {author} {\bibinfo {author} {\bibfnamefont {H.~W.}\ \bibnamefont
  {Weber}},\ }\href@noop {} {\bibfield  {journal} {\bibinfo  {journal}
  {International Journal of Modern Physics E}\ }\textbf {\bibinfo {volume}
  {20}},\ \bibinfo {pages} {1325} (\bibinfo {year} {2011})}\BibitemShut
  {NoStop}%
\bibitem [{\citenamefont {Fl{\"u}kiger}\ \emph {et~al.}(2013)\citenamefont
  {Fl{\"u}kiger}, \citenamefont {Baumgartner}, \citenamefont {Eisterer},
  \citenamefont {Weber}, \citenamefont {Spina}, \citenamefont {Scheuerlein},
  \citenamefont {Senatore}, \citenamefont {Ballarino},\ and\ \citenamefont
  {Bottura}}]{Fluekiger:RRP_PIT_irrad}%
  \BibitemOpen
  \bibfield  {author} {\bibinfo {author} {\bibfnamefont {R.}~\bibnamefont
  {Fl{\"u}kiger}}, \bibinfo {author} {\bibfnamefont {T.}~\bibnamefont
  {Baumgartner}}, \bibinfo {author} {\bibfnamefont {M.}~\bibnamefont
  {Eisterer}}, \bibinfo {author} {\bibfnamefont {H.~W.}\ \bibnamefont {Weber}},
  \bibinfo {author} {\bibfnamefont {T.}~\bibnamefont {Spina}}, \bibinfo
  {author} {\bibfnamefont {C.}~\bibnamefont {Scheuerlein}}, \bibinfo {author}
  {\bibfnamefont {C.}~\bibnamefont {Senatore}}, \bibinfo {author}
  {\bibfnamefont {A.}~\bibnamefont {Ballarino}}, \ and\ \bibinfo {author}
  {\bibfnamefont {L.}~\bibnamefont {Bottura}},\ }\href@noop {} {\bibfield
  {journal} {\bibinfo  {journal} {IEEE Transactions on Applied
  Superconductivity}\ }\textbf {\bibinfo {volume} {23}},\ \bibinfo {pages}
  {8001404 (4 pp.)} (\bibinfo {year} {2013})}\BibitemShut {NoStop}%
\bibitem [{\citenamefont {Ekin}(2010)}]{Ekin:USL}%
  \BibitemOpen
  \bibfield  {author} {\bibinfo {author} {\bibfnamefont {J.~W.}\ \bibnamefont
  {Ekin}},\ }\href@noop {} {\bibfield  {journal} {\bibinfo  {journal}
  {Superconductor Science and Technology}\ }\textbf {\bibinfo {volume} {23}},\
  \bibinfo {pages} {1} (\bibinfo {year} {2010})}\BibitemShut {NoStop}%
\bibitem [{\citenamefont {Baumgartner}\ \emph {et~al.}(2012)\citenamefont
  {Baumgartner}, \citenamefont {Eisterer}, \citenamefont {Weber}, \citenamefont
  {Fl{\"u}kiger}, \citenamefont {Bordini}, \citenamefont {Bottura},\ and\
  \citenamefont {Scheuerlein}}]{Baumgartner:Jc_mag}%
  \BibitemOpen
  \bibfield  {author} {\bibinfo {author} {\bibfnamefont {T.}~\bibnamefont
  {Baumgartner}}, \bibinfo {author} {\bibfnamefont {M.}~\bibnamefont
  {Eisterer}}, \bibinfo {author} {\bibfnamefont {H.~W.}\ \bibnamefont {Weber}},
  \bibinfo {author} {\bibfnamefont {R.}~\bibnamefont {Fl{\"u}kiger}}, \bibinfo
  {author} {\bibfnamefont {B.}~\bibnamefont {Bordini}}, \bibinfo {author}
  {\bibfnamefont {L.}~\bibnamefont {Bottura}}, \ and\ \bibinfo {author}
  {\bibfnamefont {C.}~\bibnamefont {Scheuerlein}},\ }\href@noop {} {\bibfield
  {journal} {\bibinfo  {journal} {IEEE Transactions on Applied
  Superconductivity}\ }\textbf {\bibinfo {volume} {22}},\ \bibinfo {pages}
  {6000604 (4 pp.)} (\bibinfo {year} {2012})}\BibitemShut {NoStop}%
\bibitem [{\citenamefont {Baumgartner}(2013)}]{Baumgartner:PhD}%
  \BibitemOpen
  \bibfield  {author} {\bibinfo {author} {\bibfnamefont {T.}~\bibnamefont
  {Baumgartner}},\ }\emph {\bibinfo {title} {Effects of Fast Neutron
  Irradiation on Critical Currents and Intrinsic Properties of State-of-the-Art
  {Nb}$_3${Sn} Wires}},\ \href@noop {} {Ph.D. thesis},\ \bibinfo  {school}
  {Vienna University of Technology} (\bibinfo {year} {2013}),\ \bibinfo {note}
  {\url{http://www.ub.tuwien.ac.at/diss/AC10774721.pdf}}\BibitemShut {NoStop}%
\bibitem [{\citenamefont {Dew-Hughes}(1974)}]{Dew-Hughes:flux_pinning}%
  \BibitemOpen
  \bibfield  {author} {\bibinfo {author} {\bibfnamefont {D.}~\bibnamefont
  {Dew-Hughes}},\ }\href@noop {} {\bibfield  {journal} {\bibinfo  {journal}
  {Philosophical Magazine}\ }\textbf {\bibinfo {volume} {30}},\ \bibinfo
  {pages} {293} (\bibinfo {year} {1974})}\BibitemShut {NoStop}%
\bibitem [{\citenamefont {Fl{\"u}kiger}\ \emph
  {et~al.}(2008{\natexlab{b}})\citenamefont {Fl{\"u}kiger}, \citenamefont
  {Senatore}, \citenamefont {Cesaretti}, \citenamefont {Buta}, \citenamefont
  {Uglietti},\ and\ \citenamefont {Seeber}}]{Fluekiger:optimization}%
  \BibitemOpen
  \bibfield  {author} {\bibinfo {author} {\bibfnamefont {R.}~\bibnamefont
  {Fl{\"u}kiger}}, \bibinfo {author} {\bibfnamefont {C.}~\bibnamefont
  {Senatore}}, \bibinfo {author} {\bibfnamefont {M.}~\bibnamefont {Cesaretti}},
  \bibinfo {author} {\bibfnamefont {F.}~\bibnamefont {Buta}}, \bibinfo {author}
  {\bibfnamefont {D.}~\bibnamefont {Uglietti}}, \ and\ \bibinfo {author}
  {\bibfnamefont {B.}~\bibnamefont {Seeber}},\ }\href@noop {} {\bibfield
  {journal} {\bibinfo  {journal} {Superconductor Science and Technology}\
  }\textbf {\bibinfo {volume} {21}},\ \bibinfo {pages} {054015 (8 pp.)}
  (\bibinfo {year} {2008}{\natexlab{b}})}\BibitemShut {NoStop}%
\bibitem [{\citenamefont {Rodrigues~{Jr.}}\ \emph {et~al.}(1995)\citenamefont
  {Rodrigues~{Jr.}}, \citenamefont {Thieme}, \citenamefont {Pinatti},\ and\
  \citenamefont {Foner}}]{Rodrigues:grain_boundaries}%
  \BibitemOpen
  \bibfield  {author} {\bibinfo {author} {\bibfnamefont {D.}~\bibnamefont
  {Rodrigues~{Jr.}}}, \bibinfo {author} {\bibfnamefont {C.~L.~H.}\ \bibnamefont
  {Thieme}}, \bibinfo {author} {\bibfnamefont {D.~G.}\ \bibnamefont {Pinatti}},
  \ and\ \bibinfo {author} {\bibfnamefont {S.}~\bibnamefont {Foner}},\
  }\href@noop {} {\bibfield  {journal} {\bibinfo  {journal} {IEEE Transactions
  on Applied Superconductivity}\ }\textbf {\bibinfo {volume} {5}},\ \bibinfo
  {pages} {1607} (\bibinfo {year} {1995})}\BibitemShut {NoStop}%
\bibitem [{\citenamefont {Seibt}(1975)}]{Seibt:deuteron_irrad}%
  \BibitemOpen
  \bibfield  {author} {\bibinfo {author} {\bibfnamefont {E.}~\bibnamefont
  {Seibt}},\ }\href@noop {} {\bibfield  {journal} {\bibinfo  {journal} {IEEE
  Transactions on Magnetics}\ }\textbf {\bibinfo {volume} {11}},\ \bibinfo
  {pages} {174} (\bibinfo {year} {1975})}\BibitemShut {NoStop}%
\bibitem [{\citenamefont {K{\"u}pfer}\ \emph {et~al.}(1980)\citenamefont
  {K{\"u}pfer}, \citenamefont {Meier-Hirmer},\ and\ \citenamefont
  {Reichert}}]{Kuepfer:Jc_irrad}%
  \BibitemOpen
  \bibfield  {author} {\bibinfo {author} {\bibfnamefont {H.}~\bibnamefont
  {K{\"u}pfer}}, \bibinfo {author} {\bibfnamefont {R.}~\bibnamefont
  {Meier-Hirmer}}, \ and\ \bibinfo {author} {\bibfnamefont {T.}~\bibnamefont
  {Reichert}},\ }\href@noop {} {\bibfield  {journal} {\bibinfo  {journal}
  {Journal of Applied Physics}\ }\textbf {\bibinfo {volume} {51}},\ \bibinfo
  {pages} {1121} (\bibinfo {year} {1980})}\BibitemShut {NoStop}%
\bibitem [{\citenamefont {Maier}\ and\ \citenamefont
  {Seibt}(1981)}]{Maier:irradiated_Nb3Sn}%
  \BibitemOpen
  \bibfield  {author} {\bibinfo {author} {\bibfnamefont {P.}~\bibnamefont
  {Maier}}\ and\ \bibinfo {author} {\bibfnamefont {E.}~\bibnamefont {Seibt}},\
  }\href@noop {} {\bibfield  {journal} {\bibinfo  {journal} {Applied Physics
  Letters}\ }\textbf {\bibinfo {volume} {39}},\ \bibinfo {pages} {175}
  (\bibinfo {year} {1981})}\BibitemShut {NoStop}%
\bibitem [{\citenamefont {F{\"a}hnle}(1977)}]{Faehnle:influence_1}%
  \BibitemOpen
  \bibfield  {author} {\bibinfo {author} {\bibfnamefont {M.}~\bibnamefont
  {F{\"a}hnle}},\ }\href@noop {} {\bibfield  {journal} {\bibinfo  {journal}
  {Physica Status Solidi (B)}\ }\textbf {\bibinfo {volume} {84}},\ \bibinfo
  {pages} {245} (\bibinfo {year} {1977})}\BibitemShut {NoStop}%
\bibitem [{\citenamefont {Bett}(1974)}]{Bett:irrad}%
  \BibitemOpen
  \bibfield  {author} {\bibinfo {author} {\bibfnamefont {R.}~\bibnamefont
  {Bett}},\ }\href@noop {} {\bibfield  {journal} {\bibinfo  {journal}
  {Cryogenics}\ }\textbf {\bibinfo {volume} {14}},\ \bibinfo {pages} {361}
  (\bibinfo {year} {1974})}\BibitemShut {NoStop}%
\bibitem [{\citenamefont {Meier-Hirmer}\ \emph {et~al.}(1981)\citenamefont
  {Meier-Hirmer}, \citenamefont {Reichert},\ and\ \citenamefont
  {K{\"u}pfer}}]{Meier-Hirmer:V3Si_1}%
  \BibitemOpen
  \bibfield  {author} {\bibinfo {author} {\bibfnamefont {R.}~\bibnamefont
  {Meier-Hirmer}}, \bibinfo {author} {\bibfnamefont {T.}~\bibnamefont
  {Reichert}}, \ and\ \bibinfo {author} {\bibfnamefont {H.}~\bibnamefont
  {K{\"u}pfer}},\ }\href@noop {} {\bibfield  {journal} {\bibinfo  {journal}
  {IEEE Transactions on Magnetics}\ }\textbf {\bibinfo {volume} {17}},\
  \bibinfo {pages} {997} (\bibinfo {year} {1981})}\BibitemShut {NoStop}%
\bibitem [{\citenamefont {Helfand}\ and\ \citenamefont
  {Werthamer}(1966)}]{WHH:Hc2_II}%
  \BibitemOpen
  \bibfield  {author} {\bibinfo {author} {\bibfnamefont {E.}~\bibnamefont
  {Helfand}}\ and\ \bibinfo {author} {\bibfnamefont {N.~R.}\ \bibnamefont
  {Werthamer}},\ }\href@noop {} {\bibfield  {journal} {\bibinfo  {journal}
  {Physical Review}\ }\textbf {\bibinfo {volume} {147}},\ \bibinfo {pages}
  {288} (\bibinfo {year} {1966})}\BibitemShut {NoStop}%
\end{thebibliography}%

\end{document}